\documentstyle[12pt]{article}
\def\simlt{\stackrel{<}{{}_\sim}}

\def\be{\begin{equation}}
\def\ee{\end{equation}}
\def\bear{\be\begin{array}}
\def\eear{\end{array}\ee}
\def\bea{\begin{eqnarray}}
\def\eea{\end{eqnarray}}
\def\SM{Standard Model }
\def\baselinestretch{1}
\begin{document}
%%%%%%%%%%%%%%%%%%%%%%%%%%% subequations.sty %%%%%%%%%%%%%%%%%%%%%%%%
\catcode`@=11
\newtoks\@stequation
\def\subequations{\refstepcounter{equation}%
  \edef\@savedequation{\the\c@equation}%
  \@stequation=\expandafter{\theequation}%   %only want \theequation
  \edef\@savedtheequation{\the\@stequation}% % expanded once
  \edef\oldtheequation{\theequation}%
  \setcounter{equation}{0}%
  \def\theequation{\oldtheequation\alph{equation}}}
\def\endsubequations{\setcounter{equation}{\@savedequation}%
  \@stequation=\expandafter{\@savedtheequation}%
  \edef\theequation{\the\@stequation}\global\@ignoretrue

\noindent}
\catcode`@=12
%%%%%%%%%%%%%%%%%%%%%%%%%%%%%%%%%%%%%%%%%%%%%%%%%%%%%%%%%%%%%%%%%%%%%
\begin{titlepage}
\title{{\bf Improved Higgs mass stability bound
in the Standard Model and implications for
Supersymmetry}\thanks{Work partly supported
by CICYT under contract AEN94-0928, and  by the European Union under contract
No. CHRX-CT92-0004.}
}
\author{{\bf J.A. Casas},
{\bf J.R. Espinosa}\thanks{
Supported by a grant of Comunidad de Madrid, Spain.}
and {\bf M. Quir\'os}
\\
\hspace{3cm}\\
Instituto de Estructura de la Materia, CSIC\\
Serrano 123, 28006-Madrid, Spain}
\date{}
\maketitle
%\vspace{1.5cm}
\def\baselinestretch{1.15}
\begin{abstract}
\noindent
We re-examine the lower bound on the mass of the Higgs boson, $M_H$,
from Standard Model vacuum stability including next-to-leading-log
radiative corrections. This amounts to work with the full one-loop
effective potential, $V(\phi)$, improved by two-loop RGE, and allows to
keep control of the scale invariance of $V$ in a wide range of
the $\phi$-field. Our results show that the bound is ${\cal O}\ (10\
GeV)$ less stringent than in previous estimates. In addition we
perform a detailed comparison between the SM lower
bounds on $M_H$ and the supersymmetric upper bounds on it. It turns
out that depending on the actual value of the top mass, $M_t$, the eventually
measured Higgs mass can discard the pure SM, the Minimal
Supersymmetric Standard Model or both.

\end{abstract}

\thispagestyle{empty}

\leftline{}
%\leftline{CERN--TH.7334/94}
\leftline{September 1994}
\leftline{}

\vskip-18.cm
\rightline{}
\rightline{{\bf IEM--FT--93/94}}
\rightline{{\bf hep-ph/9409458}}
\vskip3in

\end{titlepage}
\newpage
%%%%%%%%%%%%%%%%%%%%%%%%%%%%%%%%%%%%%%%%%%%%%%%%%%%%%%%%%%%%%%%%%%%
\setcounter{page}{1}

\section{Introduction}

The vacuum stability requirement in the \SM (SM)
imposes a severe lower bound on the mass of the Higgs boson, $M_H$ [1-4],
which depends on the mass of the top quark $M_t$ and on the cut-off
$\Lambda$ beyond which the SM is no longer valid. Roughly speaking, this is
due to the fact that the top Yukawa coupling, $h_t$, drives the quartic
coupling of the Higgs potential, $\lambda$, to negative values at
large scales, thus destabilizing the standard electroweak vacuum.

This bound is very relevant for us since it lies in a region that could be
accesible for present (LEP, Tevatron) and future (LEP-200, LHC) accelerators.
It is therefore worthwile to re-examine this important issue taking
into account recent
improvements on the evaluation of the effective potential and the
corresponding Higgs mass \cite{Einhorn,CEQR}. This is the aim of this
article \footnote{The results contained in
this article have been presented by one of us in {\em Physics from Planck
scale to electroweak scale}, Warsaw, Poland, 21--24 September 1994
\cite{mq}.}.

In previous works, the stability
bound was obtained from the tree level potential,
improved by one-loop or two-loop renormalization group equations (RGE) for
the $\beta$- and $\gamma$- functions of the running couplings, masses and the
$\phi$-field (see e.g. refs.~\cite{LSZ,Sher}). However, it has
been shown
that the one-loop corrections to the Higgs potential are important in order
to fix the boundary conditions for the electroweak breaking and calculate
the Higgs mass in a consistent and scale-independent way. As we will see,
they are also significant to properly understand the whole structure of the
potential. In section~2 we write the one-loop corrected Higgs potential,
giving the general conditions for extremals. In section~3 we apply
the previous results to the determination of the realistic SM vacuum
(which implies a relationship between the boundary conditions of the
SM parameters) and the physical Higgs mass up to next-to-leading-log
order. For a given top mass, there is a one-to-one correspondence
between $M_H$ and the boundary condition for $\lambda(t)$. We discuss
also in this section the differences between our approach and
previous calculations. In section 4 we study the form of the Higgs
potential for large $\phi$ and determine under which circumstances it
develops a maximum and an additional minimum. This allows us to
define the conditions for vacuum stability in
the SM, from which we extract the lower bound on $M_H$ as a function
of $M_t$ and $\Lambda$. In section 5 we present our complete numerical
results. Typically we find that the lower bound on $M_H$ is ${\cal
O}(10\ GeV)$ lower than in the last previous estimate \cite{Sher}.
In addition, we perform a detailed comparison between the SM lower
bounds on $M_H$ and the supersymmetric upper bounds on it. It turns
out that depending on the actual value of $M_t$, the eventually
measured Higgs mass can discard the pure SM, the Minimal
Supersymmetric Standard Model or both.

\section{The effective potential of the \SM}

The renormalization group improved effective potential of the SM, $V$, can
be written in the 't Hooft-Landau gauge and the $\overline{MS}$ scheme as
\cite{Einhorn}
\be
\label{veff}
V[\mu(t),\lambda_i(t);\phi(t)]\equiv V_0 + V_1 + \cdots \;\; ,
\ee
where $\lambda_i\equiv (g,g',\lambda,h_t,m^2)$ runs over all
dimensionless and dimensionful couplings and $V_0$, $V_1$ are respectively
the tree level potential and the one-loop correction, namely
\subequations{
\be
\label{v0}
V_0=-{\displaystyle\frac{1}{2}}m^2(t)\phi^2(t) +
{\displaystyle\frac{1}{8}}\lambda(t)\phi^4(t),
\ee
\be
\label{v1}
V_1={\displaystyle\sum_{i=1}^5}{\displaystyle\frac{n_i}{64\pi^2}}
M_i^4(\phi)\left[\log{\displaystyle\frac{M_i^2(\phi)}{\mu^2(t)}}
-c_i\right]+\Omega(t),
\ee}
\endsubequations
with
\be
M_i^2(\phi)=\kappa_i\phi^2(t)-\kappa_i',
\ee
and
\bear{llll}
n_1=6\ ,&  \kappa_1={\displaystyle\frac{1}{4}}g^2(t)\ ,& \kappa_1'=0 \ ,&
c_1={\displaystyle\frac{5}{6}}\ ;\vspace{0.3cm}\\
n_2=3\ ,&  \kappa_2={\displaystyle\frac{1}{4}}[g^2(t)+g'^2(t)]\ ,&
\kappa_2'=0\ ,& c_2={\displaystyle\frac{5}{6}}\ ;\vspace{0.3cm}\\
n_3=-12\ ,& \kappa_3={\displaystyle\frac{1}{2}}h_t^2(t)\ ,& \kappa_3'=0\ ,&
c_3={\displaystyle\frac{3}{2}}\ ;\vspace{0.3cm}\\
n_4=1\ ,&  \kappa_4={\displaystyle\frac{3}{2}}\lambda(t)\ ,
& \kappa_4'=m^2(t)\ ,&
c_4={\displaystyle\frac{3}{2}}\ ;\vspace{0.3cm}\\
n_5=3\ ,&  \kappa_5={\displaystyle\frac{1}{2}}\lambda(t)\ ,
& \kappa_5'=m^2(t)\ ,&
c_5={\displaystyle\frac{3}{2}}\ .\vspace{0.3cm}
\eear
$M_i^2(\phi,t)$ are the tree-level expressions for the masses of the
particles that enter in the one-loop radiative corrections, namely
$M_1\equiv m_W$, $M_2\equiv m_Z$, $M_3\equiv m_t$, $M_4\equiv
m_{\rm Higgs}$, $M_5\equiv m_{\rm Goldstone}$.
$\Omega(t)\equiv \Omega[\lambda_i(t),\mu(t)]$ is
the one-loop contribution to the cosmological constant \cite{Einhorn},
which will turn to be irrelevant in our calculation.

In the previous expressions the parameters $\lambda(t)$ and $m(t)$ are the SM
quartic coupling and mass, whereas $g(t)$, $g'(t)$, $h_t(t)$ are the SU(2),
U(1) and top Yukawa couplings respectively. All of them are running with
the RGE. The running of the Higgs field is
\be
\label{field}
\phi(t)=\xi(t) \phi_c,
\ee
$\phi_c$ being the classical field and
$\xi(t)=\exp\{-\int^t_0 \gamma(t')dt'\}$,
where $\gamma(t)$ is the Higgs field anomalous dimension.
Finally the scale $\mu(t)$ is related to the
running parameter $t$ by
\be
\label{mu}
\mu(t)=\mu e^t,
\ee
where $\mu$ is a fixed scale, that we will take equal to the physical
$Z$ mass, $M_Z$.

It has been shown \cite{Lpot} that the L-loop  effective potential
improved by (L+1)-loop RGE resums all Lth-to-leading logarithm contributions.
Consequently, we will consider all the $\beta$- and $\gamma$-functions of
the previous parameters to two-loop order, so that our calculation will
be valid up to next-to-leading logarithm approximation.

As has been pointed out in ref.~\cite{Einhorn}, working with $\partial V/
\partial \phi$ (and higher derivatives) rather than with $V$ itself allows
to ignore the cosmological constant\footnote{This holds even
if we choose $\mu(t)$ to be a function of the $\phi$-field since the scale
invariant properties of $V$ allow to perform the substitution either
before or after taking the derivative $\partial/\partial\phi$ \cite{CEQR}.
We will turn to this point below.} term $\Omega$. In fact,
the structure of the potential
can be well established once we have determined the values of $\phi$, say
$\phi_{\rm ext}$, in which $V$ has extremals (maxima or minima), thus we only
need to evaluate $\partial V/
\partial \phi$ and $\partial^2 V/
\partial \phi^2$. From eq. (\ref{veff}), the value of $\phi_{\rm ext}$
is given by
\be
\label{first}
\left.\frac{\partial V}{\partial \phi(t)}\right|_{\phi(t)=
\phi_{\rm ext}(t)}=0
\;\;\Rightarrow\;\; \phi^2_{\rm ext}(t)=\frac{2m^2
+{\displaystyle\sum_i\frac{n_i\kappa_i\kappa'_i}{8\pi^2}}
\left[\log\frac{M_i^2(\phi)}{\mu^2(t)} -c_i+\frac{1}{2}
\right]}{\lambda+{\displaystyle\sum_i\frac{n_i\kappa_i^2}{8\pi^2}}
\left[\log\frac{M_i^2(\phi)}{\mu^2(t)} -c_i+\frac{1}{2}
\right]} .
\ee
The second derivative is given by
\be
\label{second}
\left.\frac{\partial^2 V}{\partial \phi^2(t)}\right|_{\phi(t)=
\phi_{\rm ext}(t)}
=2m^2+\sum_i\frac{n_i\kappa_i\kappa'_i}{8\pi^2}
\left[\log\frac{M_i^2(\phi)}{\mu^2(t)} -c_i+\frac{1}{2}
\right] +\phi^2(t)\sum_i\frac{n_i\kappa_i^2}{8\pi^2},
\ee
where we have used (\ref{first}). Eq. (\ref{second}) can be written
in a more suggestive form as
\bea
\label{secondbis}
\left.\frac{\partial^2 V}{\partial \phi^2(t)}\right|_{\phi(t)=
\phi_{\rm ext}(t)}
&=&2m^2+\frac{1}{2}(\beta_{\lambda}-4\gamma\lambda)\phi^2(t)
\vspace{0.3cm}\nonumber\\
&+&\frac{1}{2}(\beta_{m^2}-2\gamma m^2)\left[\sum_{i,\kappa_i'\neq 0}
\log\frac{M_i^2(\phi)}{\mu^2(t)} -2 \right] ,
\eea
where $\beta_{\lambda}$, $\beta_{m^2}$ and $\gamma$ are one-loop
$\beta$- and $\gamma$-functions:
\bea
\label{betas}
16\pi^2\beta_{\lambda}&=&12\left(\lambda^2-h^4_t+\lambda h_t^2\right) -
\left(3g'^2+9g^2\right) \lambda
+{\displaystyle\frac{9}{4}}\left[{\displaystyle\frac{1}{3}}g'^4
+{\displaystyle\frac{2}{3}}g'^2g^2+g^4\right],
\vspace{0.2cm}\nonumber\\
16\pi^2\beta_{m^2}&=&m^2\left[6\lambda+6
h_t^2-{\displaystyle\frac{9}{2}}g^2-
{\displaystyle\frac{3}{2}}g'^2\right],
\vspace{0.2cm}\nonumber\\
16\pi^2\gamma&=&3\,\left(h^2_t-
{\displaystyle\frac{1}{4}}g'^2-
{\displaystyle\frac{3}{4}}g^2\right).
\eea

The form of eq. (\ref{secondbis}) reflects the scale invariance of $V$.
Indeed, it is straightforward to check that the requirement
$d V/dt=0$ applied to the one-loop effective potential
of eq. (\ref{veff}) at any value of $\phi$ implies
\subequations{
\be
{\displaystyle\frac{1}{8}}\beta_{\lambda}-{\displaystyle\frac{1}{2}}
\gamma\lambda=
{\displaystyle\sum_i}{\displaystyle\frac{n_i\kappa_i^2}{32\pi^2}},
\label{p4inv}
%\vspace{0.3cm}
\ee
\be
{\displaystyle\frac{1}{2}}\beta_{m^2}-\gamma m^2=
{\displaystyle\sum_i}{\displaystyle\frac{n_i\kappa_i\kappa_i'}{16\pi^2}},
\label{p2inv}
%\vspace{0.3cm}
\ee
\be
{\displaystyle\sum_i}{\displaystyle\frac{n_i\kappa_i'^2}{32\pi^2}}=
{\displaystyle\frac{\partial\Omega}{\partial t}}\label{p0inv},
\ee}
\endsubequations
up to two-loop corrections. Substituting (\ref{p4inv}) and (\ref{p2inv})
in (\ref{second}) we obtain eq. (\ref{secondbis}).

To conclude this section, we would like to stress the fact that even though
the whole effective potential is scale-invariant, the one-loop approximation
is {\em not}. Therefore, one needs a criterion to choose the appropriate
renormalization scale in the previous equations. As was shown in \cite{CEQR},
a sensible criterion is to choose the scale, say $\mu^*=\mu(t^*)$, where the
effective potential is more scale-independent. $\mu^*$ turns out to be a
certain average of the $M_i(\phi)$ masses. Actually, it was shown in
\cite{CEQR}, and we would like to emphasize it here, that any choice of $\mu^*$
around the optimal value produces the same results for physical quantities up
to tiny differences.

\section{The realistic minimum and the Higgs mass}

We follow in this section the approach in ref.~\cite{CEQR}, to which we
refer the reader for further details.
The boundary conditions for the parameters of the Higgs potential,
$m^2(t)$, $\lambda(t)$, are constrained by the fact that $V$ must develop an
electroweak breaking minimum consistent with the experimental observations.
In the framework of the Standard Model this ``realistic" minimum corresponds
to our present vacuum. Accordingly, we must impose
\be
\label{vev}
\langle\phi(t_Z)\rangle=v=(\sqrt{2}G_{\mu})^{-1/2}=246.22\ {\rm GeV},
\ee
where $t_Z$ is defined as $\mu(t_Z)=M_Z$ and $v$
is the ``measured" VEV for the Higgs field\footnote{We are
neglecting here one-loop electroweak radiative corrections
to the muon $\beta$-decay slightly
modifying the relation (\ref{vev}), see e.g.
ref.~\cite{peris}.} ~\cite{ara}. On the other hand
$\langle\phi(t)\rangle$ must be obtained from the minimum
condition (\ref{first}). If we evaluate eq. (\ref{first}) at the convenient
scale $\mu^*=\mu(t^*)$, what we get is  $\langle\phi(t^*)\rangle$
rather than $\langle\phi(t_Z)\rangle$. Then, $\langle\phi(t_Z)\rangle$ is
obtained through its RGE
\be
\label{vevstar1}
\langle\phi(t_Z)\rangle=\langle\phi(t^*)\rangle
{\displaystyle\frac{\xi(t_Z)}{\xi(t^*)}}.
\ee
The criterion to choose $\mu^*$ has been stated in the previous section.

We can trade $\langle\phi(t^*)\rangle$ by $m^2(t^*)$ from
the condition of minimum (\ref{first}),
which translates, using (\ref{vevstar1}) and (\ref{vev}), into the boundary
condition for $m^2(t)$:
\bear{cc}
\label{bcm}
{\displaystyle\frac{m^2(t^*)}{v^2}}&=
{\displaystyle\frac{1}{2}}\lambda(t^*)\xi^2(t^*)
+ {\displaystyle\frac{3}{64\pi^2}}\xi^2(t^*)\left\{
{\displaystyle\frac{1}{2}}g^4(t^*)\left[\log{
\displaystyle\frac{g^2(t^*)\xi^2(t^*)v^2}{4\mu^2(t^*)}}
-{\displaystyle\frac{1}{3}}\right]\right.\vspace{.5cm}\\
&+{\displaystyle\frac{1}{4}}\left[g^2(t^*)+g'^2(t^*)\right]^2
\left[\log{\displaystyle
\frac{\left[g^2(t^*)+g'^2(t^*)\right]\xi^2(t^*)v^2}{4\mu^2(t^*)}}
-{\displaystyle\frac{1}{3}}\right]\vspace{.5cm}\\
&-\left.4h_t^4(t^*)\left[\log{
\displaystyle\frac{h_t^2(t^*)\xi^2(t^*)v^2}{2\mu^2(t^*)}}
-1\right]\right\},
\eear
where we have neglected the Higgs and Goldstone boson contributions.

The running Higgs mass, $m_H^2(t)$, defined as the curvature of the
scalar potential at the minimum,
can be readily obtained from (\ref{second}) evaluated at
the scale $t^*$, i.e.
\be
m_H^2(t^*)=\left.\frac{\partial^2V}{\partial\phi^2(t^*)}\right|_{\phi(t^*)
=\langle\phi(t^*)\rangle}.
\ee
The scale invariance of the second derivative of the potential,
$\frac{\partial}{\partial t}\left[\xi^2(t)\frac{\partial^2V}
{\partial\phi^2(t)}
\right]=0$,
allows us to write $m_H^2(t)$ at any arbitrary scale
\be
\label{mhrun}
m_H^2(t)=m_H^2(t^*){\displaystyle\frac{\xi^2(t^*)}{\xi^2(t)}}.
\ee
The physical (pole) Higgs mass, $M_H^2$, is then given by
\be
\label{MHphys}
M_H^2= m_{H}^2(t) +{\rm Re}[\Pi (p^2=M_H^2)-\Pi (p^2=0)],
\ee
where $\Pi(p^2)$ is the renormalized self-energy of the Higgs boson
(the $t$-dependence of (\ref{MHphys}) drops out). Explicit expressions for
$m_H^2(t^*)$ and $\Pi(M_H^2)-\Pi(0)$ can be found in ref.~\cite{CEQR}.

As has been stated above, the choice of $\mu^*$, i.e. the scale at
which we evaluate the minimum conditions, is not important for physical
quantities, provided it is within a (quite wide) region around the optimal
value. This is illustrated for the Higgs mass
with a representative example in Fig.~1
(solid line). The lack of flatness of $M_H$ reflects the effect of
all non-considered (higher order) contributions in the calculation
and, therefore, it is a measure of the total error in our estimate
of $M_H$. From Fig.~1 we can see that the error is typically
$\simlt$ 3 GeV, which is the uncertainty we should assign to our
results.
The dashed line is the corresponding result performing the
previous calculations just with the (RGE improved) tree-level part
of $V$ in eq. (\ref{veff}), as was done in refs.~\cite{LSZ,Sher}. Then,
the Higgs mass has a strong dependence on $\mu^*$. Choosing $\mu^*=M_Z$, as
it was done in refs.~\cite{LSZ,Sher},
results in an error in the estimate of $M_H$,
whose precise value depends on the top
mass and is typically of ${\cal O}(10\ GeV)$,
showing the need of a more careful treatment of the
problem, as the one exposed above.

Finally, let us note that in the previous equations the top Yukawa
coupling, $h_t(t)$, enters in several places. Therefore, the Higgs
mass depends on the boundary condition chosen for $h_t(t)$, and thus
on the top mass $M_t$. However the running top mass, defined as
$m_t(t)=vh_t(t)$, does not coincide with the physical (pole) mass,
$M_t$. In the Landau gauge the relationship between the running $m_t$ and the
physical (pole) mass $M_t$ is given by~\cite{gray}
\be
\label{mtphys}
M_t=\left\{1+{\displaystyle\frac{4}{3}}
{\displaystyle\frac{\alpha_S(M_t)}{\pi}}+
\left[16.11-1.04\sum_{i=1}^{5}\left(1-\frac{M_i}{M_t}\right)\right]
\left(\frac{\alpha_S(M_t)}{\pi}\right)^2
\right\} m_t(M_t).
\ee
where $M_i$, $i=1,\ldots,5$ represent the masses of the five lighter quarks.

Summarizing the results of this section, for a given top mass and a boundary
condition for $\lambda(t)$, the boundary condition of $m^2(t)$ is obtained
from the electroweak breaking constraint, and is given by eq.(\ref{bcm}).
Then, the Higgs mass, $M_H$, can be calculated from eq.(\ref{MHphys}).

\section{The structure of the potential for large $\phi$ and the lower
bound on $M_H$}

The structure of maxima and minima of $V$ for large $\phi$ can be studied
with the equations (\ref{first}) and (\ref{secondbis}), evaluated at a scale
$\mu(t)$ within the region where $V$ is scale-invariant. As was discussed in
\cite{CEQR} and in the previous section, $\mu(t)=\phi(t)$ is always a correct
choice (other choices, such that $\mu(t)=\phi(t)/2$, are equally valid and lead
to the same results). Then the extremal condition (\ref{first}),
neglecting the Higgs and Goldstone contributions, reads
\be
\label{phiext}
\phi^2_{\rm ext}=\frac{2m^2}{\tilde{\lambda}},
\ee
with
\bea
\label{lambdahat}
\tilde{\lambda}&=&\lambda-
{\displaystyle\frac{1}{16\pi^2}}\left\{6h_t^4\left[\log
{\displaystyle\frac{h_t^2}{2}}-1\right]-
{\displaystyle\frac{3}{4}}g^4\left[\log
{\displaystyle\frac{g^2}{4}}-{\displaystyle\frac{1}{3}}
\right]\right.\nonumber \\
&&\nonumber \\
&-&\left.{\displaystyle\frac{3}{8}}\left(g^2+g'^2\right)^2
\left[\log{\displaystyle\frac{
\left(g^2+g'^2\right)}{4}}-{\displaystyle\frac{1}{3}}\right]
\right\},
\eea
[all quantities in (\ref{phiext},\ref{lambdahat}) are evaluated at
$\mu(t)=\phi_{\rm ext}(t)$]. From (\ref{phiext}) we see that, if $V$
develops an extremal for large values of $\phi$, this must occur for
a value of $\phi$ such that
\be
\label{lext}
0<\tilde{\lambda}[\mu(t)=\phi(t)]\ll 1.
\ee
On the other hand, for large values of $\phi$ eq. (\ref{secondbis})
can be very accurately expressed as
\bea
\label{secblarge}
\left.\frac{\partial^2 V}{\partial \phi^2(t)}\right|_{\phi(t)=
\phi_{\rm ext}(t)}
&=&\frac{1}{2}(\beta_{\lambda}-4\gamma\lambda)\phi^2(t).
\eea
Since near the extremum $\tilde{\lambda}$, and thus $\lambda$, is very
small, we see from (\ref{secblarge}) that depending on the sign of
$\beta_{\lambda}$ we will have a maximum or a minimum.

We have illustrated these features in Fig.~2 with a typical example.
Fig.~2a represents the evolution of $\lambda$ (dashed line) and
$\tilde{\lambda}$ (solid line)
with $\mu(t)$. It is worth noticing that they do not cross the
horizontal axis at the same value of $\mu(t)$, but they differ by more
than one order of magnitude. This is important since the point
where the maximum of the potential is located, say $\phi_{MAX}$, does
correspond to $\tilde{\lambda}\sim 0$ rather than ${\lambda}\sim 0$ (see
eq.~(\ref{lext})). This is apparent in Fig.~2b, where the scalar
potential, $V(\phi)$, has been represented\footnote{The function
of $V$ that has been plotted in Fig.~2b has been chosen in order to give
a continuous and faithful representation of $V$ in logarithmic units.}.
The open diamond and square in Fig.~2a correspond to the position of the
maximum and the minimum, respectively,  of $V(\phi)$.
Notice also that for values of $\phi$ very
slightly higher than $\phi_{MAX}$, the potential is negative and much
deeper than the realistic minimum. This is simply because for values
of $\mu(t)$ just beyond $\mu_{MAX}=\phi_{MAX}$, the value
of $\tilde{\lambda}$ becomes negative and the
potential is dominated by the contribution
$\frac{1}{8}\tilde{\lambda}\phi^4$ (compare eqs.~(\ref{veff}) and
(\ref{lambdahat})). Consequently, a sensible criterion for a model to
be safe is to require one of the two following conditions:
\begin{description}
\item[{\it a)}] The potential has no maximum.

\item[{\it b)}] The maximum occurs for $\phi_M>\Lambda$,
\end{description}
\noindent
where we recall that $\Lambda$ is the cut-off beyond which the \SM
is no longer valid. In the following we will assume $\Lambda\leq
10^{19}$ GeV. With this criterion [in particular condition ({\it b})]
we see that the model represented in Fig.~2 is acceptable for
$\Lambda\leq 2.7\times 10^{11}$ GeV.
Beyond this scale the stability of the vacuum requires the appearance
of new physics. Note from this discussion that conditions
({\it a}), ({\it b}) are not equivalent to require
$\lambda(\mu)>0$ for $\mu(t)<\Lambda$, as is usually done.
Instead, the significant parameter is $\tilde{\lambda}$ rather than
${\lambda}$.

We have represented in Fig.~3 the evolution of
$\tilde{\lambda}$ for a set of models with the same value of $M_t$ as the
previous one ($M_t=160$ GeV), but with different boundary conditions
for $\lambda(t)$, and thus different values of $M_H$ (obtained as
explained in sect.~3). The thickest line ($M_H=101$ GeV), which is
tangent to the horizontal axis at
$\Lambda_0=4\times 10^{11}$ GeV, represents a
limiting case: for $M_H> 101$ GeV the potential has no maximum and
therefore is completely safe for any choice of $\Lambda$ [see
condition ({\em a}) above]; for $M_H<101$ GeV the models are safe
depending on the value of $\Lambda<\Lambda_0$.
Hence, for $\Lambda>\Lambda_0$ the lower bound
on $M_H$ is insensitive to the value of $\Lambda$. The situation
depicted in this example, i.e. for $M_t=160$ GeV, typically
occurs when the top mass is rather small, since then the top Yukawa
coupling is too small at large scales to maintain $\beta_{\lambda}$
negative [see eq. (\ref{betas})]. When the top mass is higher the
situation is different and is illustrated with the case $M_t=174$ GeV
in Fig.~4. There we see that for each value of $M_H$,
$\tilde{\lambda}(t)$ crosses the horizontal
axis at most once, and there is a
value of $M_H$ for which the cross occurs at $\Lambda=10^{19}$ GeV.
In consequence, for $\Lambda\leq10^{19}$ GeV there is a one-to-one
correspondence between the choice of $\Lambda$ and the value of the
lower bound on $M_H$.

Finally, we would like to point out that generically the potential is
{\em not} unbounded from below since whenever there is a maximum,
there is an additional minimum for a larger value of $\phi$,
as illustrated in Fig.~2b. This is because $\tilde{\lambda}$ gets
back to the positive rage for large enough scales. Hence,
the potential becomes eventually positive and monotonically
increasing, although in some cases,
e.g. for $M_t=174$ GeV, this occurs for values of $\phi$ beyond
$10^{19}$ GeV.
\section{Numerical results and comparison with SUSY bounds}

As stated in the Introduction and has become clear in previous
sections, the lower bound on $M_H$ is a function of $M_t$ and
$\Lambda$. However,
apart from the previously estimated error $\simlt$ 3 GeV
in our calculation, there is an additional source of uncertainty
coming from the value of $\alpha_S$, which enters in several places
in the previous calculation. The most recent estimate of $\alpha_S$
gives
\cite{Langa}
\be
\label{alphas}
\alpha_S=0.124\pm0.006.
\ee
Using the central value of (\ref{alphas}), we have represented in
Fig.~5 the lower bound on $M_H$ as a function of $M_t$ for different
values of $\Lambda$. The form of the curves is easily understandable
from the discussion of the previous section. In Fig.~6,
which corresponds to the pure SM case, we have fixed
$\Lambda$ at its maximum value, $\Lambda=10^{19}$ GeV, and
represented the lower bound on $M_H$ for the central value of
$\alpha_S$ in (\ref{alphas}) (diagonal solid line) and the two
extreme values (diagonal dashed lines).

If we use the recent evidence for the top quark
production at CDF with a mass $M_t=174\pm 17$ GeV
\cite{cdf}, we obtain the following lower bound on $M_H$:
\be
\label{174bound}
M_H> 128\pm 33\ GeV,
\ee
i.e. $M_H>$ 95 GeV (1$\sigma$).
If the Higgs is observed in the present or forthcoming accelerators
with a mass below the bound of eq. (\ref{174bound}), this would be a
clear signal of new physics beyond the \SM .

Comparing these bounds with the last evaluation performed in
ref.~\cite{Sher}, we see that our values of $M_H$ are
lower than in \cite{Sher} by an amount which increases with $M_t$,
going from $\sim 5$ GeV for $M_t\sim 130$ GeV, to $\sim 15$ GeV for
$M_t \sim 200$ GeV.
As has been discussed in sect.~3,
the main reason of this difference is the way in which the Higgs mass
was computed in ref.~\cite{Sher}. Accordingly, our results give more
room to the Higgs mass in the framework of the Standard Model.

It is very interesting to perform a comparison between the SM
{\em lower} bounds on $M_H$ previously obtained and the
supersymmetric {\em upper} bounds on $M_H$ \cite{susy,CEQR}
\footnote{Previous comparisons based on old results can be found in
ref.~\cite{discr}}. We
briefly recall that in the Minimal Supersymmetric Standard Model
(MSSM) the Higgs quartic coupling is not a free parameter, but is
given by a certain combination of the $g^2$, $g'^2$ gauge couplings.
Thus, experimental data constrain the boundary condition for
$\lambda$, which cannot be as large as we want, leading to strong
upper bounds on $M_H$\footnote{It is interesting to note that for the
same reasons, the value of $\lambda$ at large scales remains always
positive and hence there are no corresponding
lower bounds on $M_H$ from vacuum
stability in the MSSM.}. These bounds depend on three parameters
(besides $M_t$): $\Lambda_{S}$, i.e. the scale below which
supersymmetry (SUSY) decouples (from naturality reasons
$\Lambda_{S}\simlt 1$ TeV \cite{fine}); $\tan\beta$, i.e. the ratio
$\langle H_2 \rangle/\langle H_1 \rangle$ of the two supersymmetric
Higgs doublets; and $X_t=A_t+\mu/\tan\beta$,
i.e. the mixing between stops, which
is responsible for the threshold correction to the Higgs quartic
coupling\footnote{For more details see e.g. ref.~\cite{CEQR}.}.
The larger threshold correction and $\tan\beta$, the less stringent
the SUSY bounds. Therefore, the most conservative situation takes
place considering maximum threshold correction (which is achieved for
$X_t^2=6\Lambda_{S}^2$) and $\tan\beta=\infty$. Likewise, the larger
$\Lambda_{S}$, the less stringent the bounds; but, as mentioned
above, it is not sensible to consider $\Lambda_{S}$ much larger
than $1$ TeV. Consequently, to be in the safe side, we have
represented in Fig.~6 the MSSM upper bounds (transverse solid and
dashed lines), as recently obtained up
to next-to-leading-log order in ref.~\cite{CEQR}, in the most
conservative situation with $\Lambda_{S}=1$ TeV.

Of course, since in the SUSY case $\Lambda_{S}$ indicates the
appearance of new physics, the supersymmetric upper bounds must be
consistent with the SM lower bounds for $\Lambda=\Lambda_{S}$, as
in fact they are. However, setting $\Lambda=10^{19}$ GeV,
i.e. the pure SM case, as in
Fig.~6, we find that this is not so for all the top masses, leading
to an interesting situation. We can distinguish three zones in
Fig.~6:
\begin{description}
\item[{\it i)}] For $M_t=173\pm 4$ GeV, i.e.
the crossing area of the SM
and MSSM curves, the eventually measured Higgs mass will be
compatible either with the pure SM  or with the MSSM, but not with
both at the same time (unless $M_H=124$ with high accuracy).
Accordingly, the experimental Higgs mass either will discard the MSSM
or will be a clear signal of new physics beyond the SM compatible
with the MSSM.

\item[{\it ii)}] For $M_t<169$  GeV, the situation is analogous, but there is a
wider range of Higgs masses (area within the two curves) compatible
with both SM and MSSM.

\item[{\it iii)}] For $M_t>177$ GeV, there is no region of
Higgs masses compatible
with the SM and MSSM simultaneously. On the contrary there is a range
of $M_H$ (within the two curves) which would discard both.
\end{description}
Perhaps the most interesting situation is ({\it i}), which occurs precisely
for a top mass in the central value of the CDF experimental estimate.

In order to facilitate the comparison between the pure SM and the
MSSM, we give below fits of the corresponding bounds on $M_H$, valid
for 150 GeV $< M_t< 200$ GeV and within $\simlt 1$ GeV of error.
\newpage
\be
\label{fits}
{\rm SM}:M_H>127.9+1.92(M_t-174)-4.25\left(
\frac{\alpha_S-0.124}{0.006}\right),
\Lambda=10^{19}\ {\rm GeV}
\nonumber
\ee
\be
{\rm MSSM}:M_H<126.1+0.75(M_t-174)-0.85\left(
\frac{\alpha_S-0.124}{0.006}\right),
\Lambda_S=10^{3}\ {\rm GeV}
\ee
where $M_H$ and $M_t$ are expressed in GeV.

%
%\section*{Acknowledgements}
%
%We thank ourselves for useful discussions and comments.

\vspace{0.4 cm}
\noindent{\em Note added}:
After this work was finished we received a preprint by G. Altarelli
and G. Isidori \cite{alt}, where the SM lower bound on $M_H$ is refined
using a different approach. Our results are typically a few GeV below
theirs, though both are consistent within the estimated errors.

%%%%%%%%%%%%%%%%%%--- References
%%%%%%%%%%%%%%%%%%%%%%%%%%%%%%%%%%%%%%%%%%%%%%%%%%%%%%%
\def\MPL #1 #2 #3 {{\em Mod.~Phys.~Lett.}~{\bf#1}\ (#2) #3 }
\def\NPB #1 #2 #3 {{\em Nucl.~Phys.}~{\bf B#1}\ (#2) #3 }
\def\PLB #1 #2 #3 {{\em Phys.~Lett.}~{\bf B#1}\ (#2) #3 }
\def\PR #1 #2 #3 {{\em Phys.~Rep.}~{\bf#1}\ (#2) #3 }
\def\PRD #1 #2 #3 {{\em Phys.~Rev.}~{\bf D#1}\ (#2) #3 }
\def\PRL #1 #2 #3 {{\em Phys.~Rev.~Lett.}~{\bf#1}\ (#2) #3 }
\def\PTP #1 #2 #3 {{\em Prog.~Theor.~Phys.}~{\bf#1}\ (#2) #3 }
\def\RMP #1 #2 #3 {{\em Rev.~Mod.~Phys.}~{\bf#1}\ (#2) #3 }
\def\ZPC #1 #2 #3 {{\em Z.~Phys.}~{\bf C#1}\ (#2) #3 }

%\newpage
\section*{Figure Captions}
\begin{description}
\item[Fig.~1] Plot of the physical Higgs mass, $M_H$, as a function
of $\mu(t^*)$ in a model with  $M_t=175$ GeV, $\Lambda=10^{19}$ GeV
and $\alpha_S=0.124$. The solid line corresponds
to our approach using the full one-loop effective potential, $V$,
improved by two-loop RGE. The dashed line is the result using only
the tree--level part of $V$, improved with two-loop RGE,
as was done in refs.~\cite{LSZ,Sher}.

\item[Fig.~2a] Plot of $\lambda$ (dashed line) and
$\tilde{\lambda}$ (solid line) as a function of the scale $\mu(t)$
for $\Lambda$ and $\alpha_S$ as in Fig.~1,
$M_t=160$ GeV and $M_H=100$ GeV.

\item[Fig.~2b] Plot of the scalar
potential, $V(\phi)$, corresponding to Fig.~2a,
represented in a convenient choice of units
as described in the text.

\item[Fig.~3] Plot of
$\tilde{\lambda}$ as a function of the scale $\mu(t)$
for $M_t=160$ GeV, $\Lambda=10^{19}$ GeV and $\alpha_S=0.124$.
The curves are shown for intervales of 5 GeV in
$M_H$ in the range $85$ GeV $\leq M_H\leq 115$ GeV. The thicker
(tangent) line corresponds to the case $M_H=101$ GeV.

\item[Fig.~4] The same as Fig.~3 for $M_t=174$ GeV and $110$ GeV
$\leq M_H\leq 140$ GeV. The thicker
line, with $\tilde{\lambda}(\mu=10^{19}\ GeV)=0$, corresponds to the
case $M_H=128$ GeV.

\item[Fig.~5] SM lower bound on $M_H$ as a function of $M_t$ for
$\alpha_S(M_Z)=0.124$ and different
values of $\Lambda$ in the range $10^{3}$ GeV $\leq \Lambda \leq
10^{19}$ GeV. The values of $\Lambda$ for consecutive curves differ
in two orders of magnitude.

\item[Fig.~6] Diagonal (thick) lines:
%the same as in Fig.~5 for
SM lower bound on $M_H$ as a function of $M_t$ for
$\Lambda=10^{19}$ GeV and $\alpha_S=0.124$ (solid line),
$\alpha_S=0.118$ (upper dashed line),
$\alpha_S=0.130$ (lower dashed line).
Transverse (thin) lines: MSSM upper bounds on $M_H$ for $\Lambda_{S}=1$ TeV
and $\alpha_S$ as in the diagonal lines.

\end{description}

\end{document}